\documentclass[preprint,12pt]{elsarticle}

\usepackage{bm}
\usepackage{amssymb}
\usepackage{mathtools}
\usepackage{xcolor}
\usepackage{hyperref}

\journal{Computer Physics Communications}

\begin{document}

\begin{frontmatter}



\title{TokaMaker: An open-source time-dependent Grad-Shafranov tool for the design and modeling of axisymmetric fusion devices}


\author[inst1]{C. Hansen\corref{cor1}}
\author[inst1]{I.G. Stewart}
\author[inst1]{D. Burgess}
\author[inst1]{M. Pharr}
\author[inst1]{S. Guizzo}
\author[inst2]{F. Logak}
\author[inst1]{A.O. Nelson}
\author[inst1]{C. Paz-Soldan}
\cortext[cor1]{Corresponding author: \texttt{christopher.hansen@columbia.edu}}
\affiliation[inst1]{organization={Applied Physics \& Applied Mathematics, Columbia University},
            city={New York},
            state={New York},
            postcode={10027},
            country={United States}}

\affiliation[inst2]{organization={École Polytechnique},
            city={Paris},
            country={France}}

\begin{abstract}
In this paper, we present a new static and time-dependent MagnetoHydroDynamic (MHD) equilibrium code, TokaMaker, for axisymmetric configurations of magnetized plasmas, based on the well-known Grad-Shafranov equation. This code utilizes finite element methods on an unstructured triangular grid to enable capturing accurate machine geometry and simple mesh generation from engineering-like descriptions of present and future devices. The new code is designed for ease of use without sacrificing capability and speed through a combination of Python, Fortran, and C/C++ components. A detailed description of the numerical methods of the code, including a novel formulation of the boundary conditions for free-boundary equilibria, and validation of the implementation of those methods using both analytic test cases and cross-code validation is shown. Results show expected convergence across tested polynomial orders for analytic and cross-code test cases.
\end{abstract}




\end{frontmatter}


\section{Introduction}
\label{sec:intro}
Fusion energy is a promising clean energy source that could help enable deep decarbonization of global energy infrastructure by providing a dispatchable carbon-free baseload electrical and thermal energy resource. The magnetic confinement approach to fusion is the closest to commercialization with several public~\cite{NACS2021,Hsu2023,STEPbook,Zhuang2019} and private~\cite{Zhang2019,Creely2020,Gryaznevich2022,Kirtley2023} efforts aimed at demonstration facilities on a decadal timescale. In order to design and operate such facilities, it is necessary to predict, control, and optimize the equilibrium plasma state where fusion reactions will occur. Most of the near-term public and commercial efforts are based on the tokamak~\cite{grad1958} configuration, which is rotationally symmetric about a central axis. For this configuration and other axisymmetric approaches, the force balance between the confined thermal energy and confining magnetic fields can be expressed using cylindrical coordinates $(R,\phi,Z)$ in terms of the Grad-Shafranov equation
\begin{equation} \label{eq:grad_shaf}
\nabla^{*} \psi = -\frac{1}{2}\frac{\partial F^2}{\partial \psi} - \mu_0 R^2 \frac{\partial P}{\partial \psi},
\end{equation}
a scalar PDE for the magnetic potential $\psi(R,Z)$. This equation is parameterized by two scalar functions of $\psi$, the thermal pressure $P$ and the radially-scaled azimuthal magnetic field $F = R B_{\phi}$ along with boundary conditions on $\psi$, which are related to externally-produced magnetic fields required to balance the hoop force and produce the desired plasma shape.

The solution of this equation has been central to the development of the tokamak, and other axisymmetric configurations, since its derivation in the late 1950s~\cite{grad1958}. As a result, many different codes have been developed over the intervening decades to solve this equation in a variety of contexts~\cite{Lao1985,Hofmann1988,Lutjens1992,corsica1997,Artaud2010,Heumann2011,Hansen2017}. However, the same long history, varied application, and relative simplicity of this equation has led to a wide and fragmented array of tools that have varying degrees of device-specific features, dependence on closed-source toolkits (eg. MATLAB), and lack of documentation that limits their portability and use by new groups.

In this paper, we describe a new tool, TokaMaker, that is designed to address these issues and provide a user-friendly, open-source Grad-Shafranov tool for the tokamak and other axisymmetric magnetic confinement concepts. The goal of this project is to provide a common tool that can be used for research, commercial, and educational applications -- providing sufficient capability and speed for all three, while also enabling easy training, use, and modification across this broad range of stakeholders. TokaMaker is part of the broader Open FUSION Toolkit, developed by the authors, which is written in a portable combination of Python, C/C++, and Fortran. The source code and pre-built binaries for Linux and macOS are publicly-available on GitHub at \href{https://github.com/hansec/OpenFUSIONToolkit}{https://github.com/hansec/OpenFUSIONToolkit}.

The remainder of the paper is structured as follows. In section~\ref{sec:num_methods}, we describe the mathematical problem for static and time-dependent applications as well as the numerical discretization and novel boundary condition formulation used in TokaMaker. Section~\ref{sec:code_desc} provides a detailed description of TokaMaker itself, including a breakdown of the different solution methods and options for its three basic modes of operation. Numerical verification tests using analytic solutions (when available) and existing prior community tools are presented in section~\ref{sec:verification}. Finally, a brief discussion and plans for future work is presented in section~\ref{sec:conclusions}.

\section{Problem description and numerical methods} \label{sec:num_methods}
\begin{figure}[h]
\begin{center}
    \includegraphics[width=0.4\textwidth]{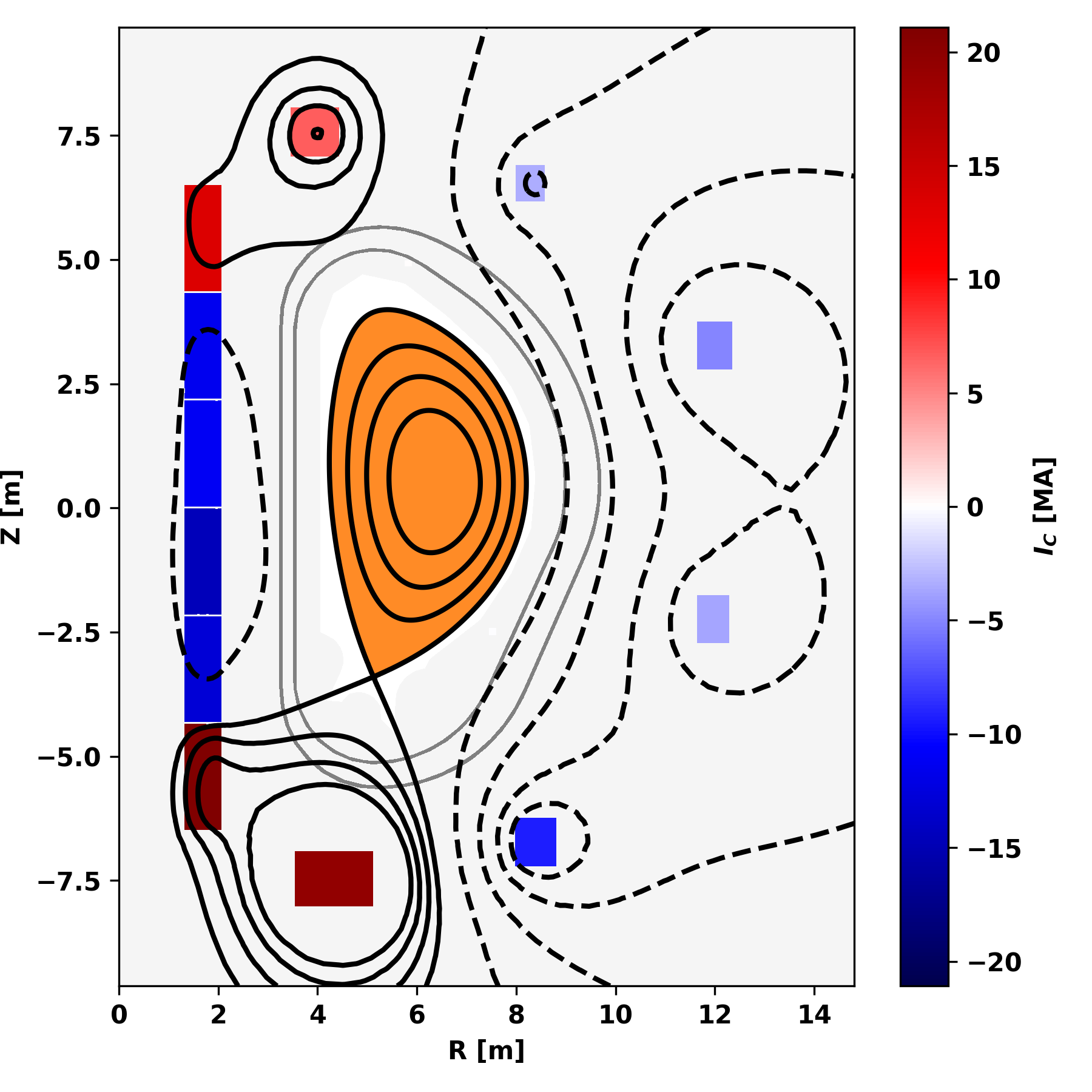} 
    \caption{Example equilibrium in ITER, showing the different regions: plasma (orange), vacuum (white and light gray), coils (color bar), and structures (dark gray) along with flux surfaces (black) for the equilibrium. Solid and dashed contours indicate values of normalized flux ($\hat{\psi}$) below and above the LCFS value respectively, where by the tokamak convention $\hat{\psi}=[0,1]$ inside the plasma.}
    \label{fig:ITER_eq_ex}
\end{center}
\end{figure}

Equation~\ref{eq:grad_shaf} describes the dependence of $\psi$ on the parameterized functions $F$ and $P$ due to MHD force balance within the plasma. However, for almost every system of interest, such a force balance also relies on currents flowing in regions outside, but near to, the plasma itself. Considering these regions, and noting that $\Delta^* \psi = -R \mu_0 J_{\phi}$, provides a broader equation to be solved
\begin{equation} \label{eq:grad_shaf_domains}
    \Delta^* \psi = 
\begin{cases}
    -\frac{1}{2}\frac{\partial F^2}{\partial \psi} - \mu_0 R^2 \frac{\partial P}{\partial \psi} & \text{if } \bm{r} \in \mathcal{P}\\
    -R \mu_0 J_{\phi} & \text{if } \bm{r} \in \mathcal{S},\mathcal{C} \\
    0 & \text{elsewhere},
\end{cases}
\end{equation}
where $\mathcal{P}$, $\mathcal{S}$, and $\mathcal{C}$ are axisymmetric domains corresponding to the plasma, passive conducting structures (eg. vacuum vessels), and coils respectively. The rest of space is treated as a vacuum, where no currents can exist. Figure~\ref{fig:ITER_eq_ex} shows an example equilibrium solution from TokaMaker for the ITER device, with each of the different regions highlighted.

Adding further detail to the description, currents flowing in the passive region $\mathcal{S}$ are driven solely by inductive voltages, while the region $\mathcal{C}$ is generally composed of multiple coils, between which the amplitude of current can vary. Further separating the system and introducing the inductive current balance $\eta \mu_0 J_{\phi} = \frac{1}{R} \frac{\partial \psi}{\partial t}$, produces the final set of equations of interest
\begin{equation} \label{eq:grad_shaf_full}
    \Delta^* \psi = 
\begin{cases}
    -\frac{1}{2}\frac{\partial F^2}{\partial \psi} - \mu_0 R^2 \frac{\partial P}{\partial \psi} & \text{if } \bm{r} \in \mathcal{P}\\
    -\frac{1}{\eta} \frac{\partial \psi}{\partial t} & \text{if } \bm{r} \in \mathcal{S}\\
    -R J_{\phi,\mathcal{C}_i} & \text{if } \bm{r} \in \mathcal{C}_i\\
    0 & \text{elsewhere},
\end{cases}
\end{equation}
where $\mathcal{C}_i$ is the $i$-th of $n$ coils in the model. The current density $J_{\phi,\mathcal{C}_i}$ can also be further expressed in terms of a fixed total current $J_{\phi,\mathcal{C}_i} = I_{\phi,\mathcal{C}_i}/\int_{\mathcal{C}_i} dA$, or an externally applied voltage $\eta \mu_0 J_{\phi} = \frac{1}{R} \frac{\partial \psi}{\partial t} + V_{ext}$, which may also include the effect of external circuits. 

\subsection{Finite element discretization} \label{sec:spatial_disc}
\begin{figure}[h]
\begin{center}
    \includegraphics[width=0.8\textwidth]{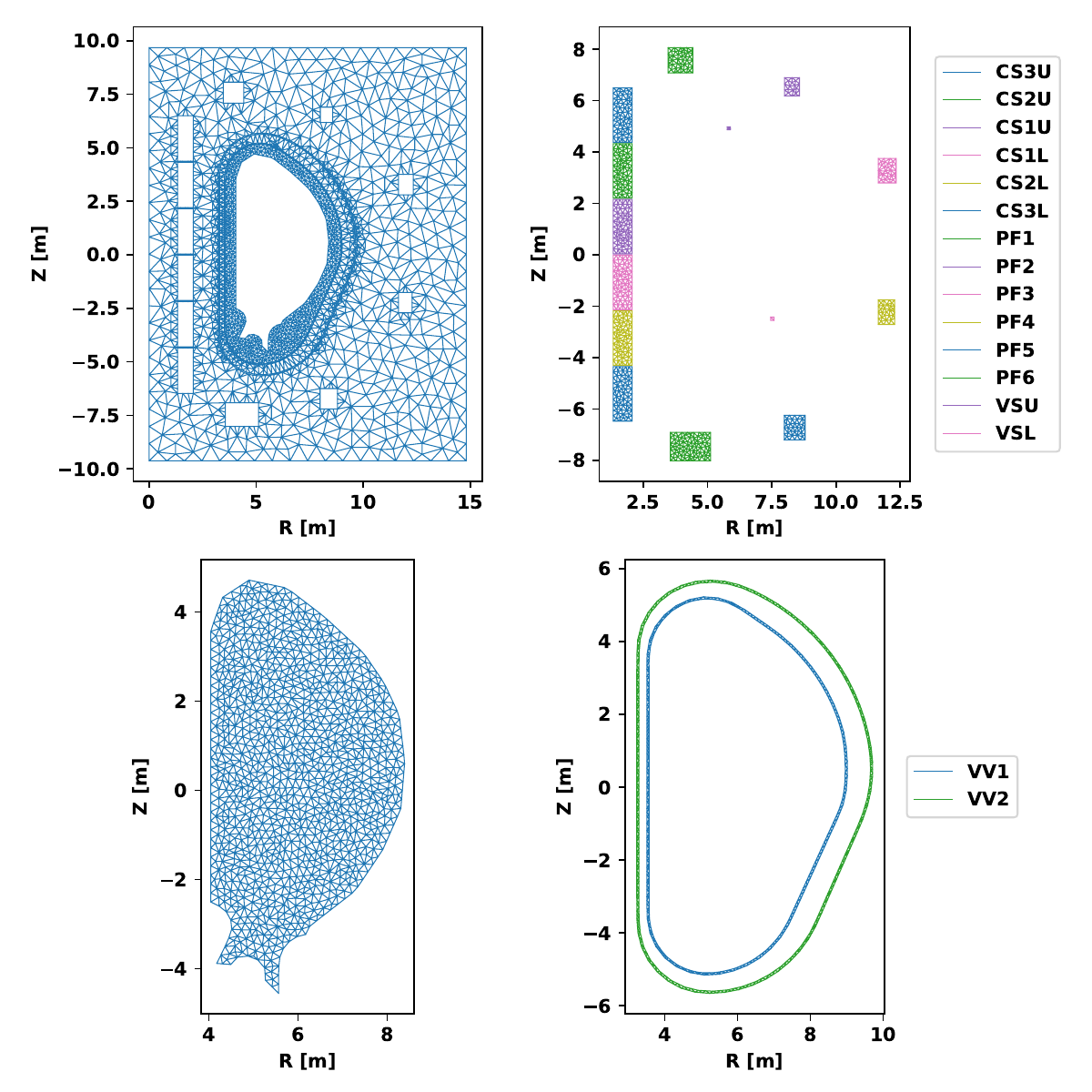} 
    \caption{Example meshes for ITER separated by region type for the vacuum (upper left), coils (upper right), structures (lower right), and plasma (lower left).}
    \label{fig:mesh_ex}
\end{center}
\end{figure}
To solve eq.~\ref{eq:grad_shaf_full}, a finite element discretization on an unstructured triangular grid is employed, similar to the approach used by the MATLAB-based FEEQS~\cite{Heumann2011,Heumann2015,Blum2019} and NICE~\cite{Faugeras2020} codes. Figure~\ref{fig:mesh_ex} shows the mesh used for the ITER equilibrium example shown in figure~\ref{fig:ITER_eq_ex}. Please note that the regions are only separated for visualization and, in reality, all regions are part of a single mesh. A $C^0$ nodal Lagrange basis set ($u$) is used with equally-spaced points with runtime-selectable polynomial degree (element order) up to four. While various spectral~\cite{Howell2014,Palha2016} and approaches with higher-order continuity (eg. $C^1$)~\cite{Jardin2004,Elarif2021} have also been proposed for the G-S equation, the $C^0$ Lagrange basis was chosen to balance numerical accuracy, geometric flexibility, and ease of implementation. For numerical convenience eq.~\ref{eq:grad_shaf_full} is first divided by $R$. Utilizing a Galerkin approach with test set ($v \in \{u\}$) and integrating by parts where appropriate yields the weak form
\begin{equation} \label{eq:grad_shaf_disc}
\begin{split}
    \int \frac{1}{R} \nabla v \cdot \nabla u \, dA &= \int_{\mathcal{P}} v \left( \frac{1}{2 R}\frac{\partial F^2}{\partial \psi} + \mu_0 R \frac{\partial P}{\partial \psi} \right) \, dA \\
    &+ \int_{\mathcal{S}} \frac{v}{\eta R} \frac{u - \psi_{t-\Delta t}}{\Delta t} \, dA + \sum_{i} \int_{\mathcal{C}_i} v J_{\phi,\mathcal{C}_i} \, dA,
\end{split}
\end{equation}
where a backward Euler method is used for $\frac{\partial \psi}{\partial t}$ and $\psi_{t-\Delta t}$ is the solution at $t-\Delta t$ (see section~\ref{sec:code_desc}). The discretized $\Delta^*$ operator, with BCs described in section~\ref{sec:boundary_conditions}, is solved directly using a sparse LU factorization package, commonly UMFPACK~\cite{Davis2004} or PARDISO~\cite{Schenk2000} through Intel's oneMKL library.

\subsection{Plasma boundary identification}
\label{sec:lcfs_finding}
While $\mathcal{S}$ and $\mathcal{C}$ have fixed geometric extents, the boundary of the plasma region ($\mathcal{P}$) is dependent on the location of the so-called Last Closed Flux Surface (LCFS). The LCFS corresponds to the first flux contour to contact the inner material structures, known as the limiter, which is shown as the boundary between white and gray regions in figure~\ref{fig:ITER_eq_ex}. As a result, the region $\mathcal{P}$ is defined by the closed level set of $\psi$ inside the limiter contour. When this closed level set contacts the limiter directly, the plasma is said to be ``limited" and the identification of $\psi_{LCFS}$ amounts to finding the maximum value of $\psi$ on the limiter contour. However, identification of $\psi_{LCFS}$ can be complicated in so-called diverted equilibria, like figure~\ref{fig:ITER_eq_ex}, where the presence of saddle points in $\psi$, which generate X-points in the level set, result in an LCFS contour that does not contact the wall directly as the point where the levelset contacts the wall is on the open side of the X-point. In this case, the location of saddles must be identified and filtered to locate the appropriate $\psi_{LCFS}$. While saddles can only exist at node points for pd=1, for pd$\geq$2 saddle points can exist anywhere within the mesh. To locate these points accurately, approximate saddles are first found using only the linear part of $\psi$ defined from mesh vertices and then refined using Newton's method (see section~\ref{sec:ver_analytic_solo}). The plasma region is then identified as the conditions $\psi \geq \psi_{LCFS}$ and $\min_{X_i} \left\{ \left(\bm{r} - \bm{r}_{X_i}\right) \cdot \left(\bm{r}_O - \bm{r}_{X_i}\right) \right\} \geq 0$, where $\bm{r}_{X_i}$ and $\bm{r}_O$ are the locations of the $i$-th X-point and O-point (magnetic axis) respectively.

\subsection{Boundary conditions}
\label{sec:boundary_conditions}
When solving eq.~\ref{eq:grad_shaf_disc}, two different boundary conditions are generally of interest: 1) The ``fixed-boundary" case where the shape of the plasma edge is specified, leading to a straightforward Dirichlet condition $\psi = C$ on the boundary, where $C$ is a pre-defined constant and 2) The ``free-boundary" case where $\psi$ at the boundary should be consistent with the vacuum projection of currents both inside and outside (eg. coils) the computational domain.

In the latter case, a simple Dirichlet condition still applies to nodes lying on the geometric axis ($\psi(R=0) = 0$). For the remaining quantities, our naive approach is to directly compute the flux from known currents,
\begin{equation} \label{eq:full_green}
\psi(\bm{r}') = \int G(\bm{r}',\bm{r}) J_{\phi} dv,
\end{equation}
where
\begin{equation} \label{eq:tor_green}
G(\bm{r}',\bm{r}) = \frac{1}{2 \pi} \sqrt{\frac{R' R}{k^2}} \left[ \left( 1-\frac{k^2}{2}\right) K(k) - E(k) \right]
\end{equation}
is the toroidal Green's function, where $E(k)$ and $K(k)$ are the complete elliptic integrals of the first and second kind respectively and $k = \frac{R' R}{\left( R' + R \right)^2 + \left( Z' - Z \right)^2}$. Equation~\ref{eq:full_green} can be expressed as a matrix coupling all current elements to each boundary node. This matrix can then be directly combined with the matrix from discretization of eq.~\ref{eq:grad_shaf} or utilized in a nested iteration approach, where the inner iteration solves the system with dirichlet boundary conditions on $\psi$ and the boundary is periodically updated using this current coupling matrix.

While this is efficient for external currents that tend to produce a tall and thin matrix (few coils), this is computationally intensive for the plasma current, which results in the coupling of all nodes in the plasma region to each node on the boundary.

A more efficient approach, that is also frequently employed by free-boundary G-S equilibrium codes, is Lackner and Von Hagenow's method~\cite{Lackner1976,JardinBook}. This method is a specific form of the virtual casing theorem~\cite{Hanson2015} that relates the magnetic field, and its vector potential, in a region to the integral of the tangential field on the surface of that region. In axisymmetric geometry this takes the form
\begin{equation} \label{eq:van_hagenow}
\psi(\bm{r}') = \int G(\bm{r}',\bm{r}) J_{\phi} dv = \oint G(\bm{r}',\bm{r}) \bm{B}_i \times \hat{\bm{n}} \cdot \hat{\bm{\phi}} dl,
\end{equation}
where $\bm{B}_i$ is a magnetic field that satisfies homogenous boundary conditions ($\bm{B}_i \cdot \hat{\bm{n}} = 0$). However, in general, the normal component of the magnetic field is non-zero on the computational boundary, so the nested iteration approach described above is employed, where eq.~\ref{eq:van_hagenow} is used in the outer loop through a separate solution of eq.~\ref{eq:grad_shaf} with homogenous boundary conditions.

In TokaMaker, this standard alternating BC has been re-formulated as a single step approach by defining the homogeneous solution in terms of the full solution and a vacuum field that nullifies $\bm{B}_i \cdot \hat{\bm{n}}$ on the boundary of the domain, which can be defined as
\begin{equation}
\bm{B}_i = \bm{B} - \oint \frac{1}{R} \nabla G(\bm{r}',\bm{r}) J_{\phi,b} \, dl,
\end{equation}
where $J_{\phi,b}$ as a surface current that yields $\psi$ on the boundary. In weak form, after rearranging matrices, this yields the following equation for the boundary nodes
\begin{equation}
0 = \mathrm{M} \mathrm{L}^{-1} \mathrm{M} \bm{\psi} + \mathrm{A} \bm{\psi} - \mathrm{A} \mathrm{P} \mathrm{L}^{-1} \mathrm{M} \bm{\psi},
\end{equation}
where $\mathrm{L} = \oint \oint v' G(\bm{r}',\bm{r}) u \, dl' dl$ and $\mathrm{M} = \oint v u \, dl$ are the boundary inductance and mass matrices respectively, $\mathrm{A} = \oint \frac{1}{R} v \nabla u \cdot \hat{\bm{n}} \, dl$ is a boundary tangential field projection matrix, and $\mathrm{P}_i = \oint G(\bm{r}'_i,\bm{r}) u \, dl$ projects the surface current to $\psi$ on a given node $i$. Note that the first and third terms are only dependent on the boundary nodes of $\psi$, where as the second term also involves interior nodes.

While integration of $\mathrm{M}$ and $\mathrm{A}$ are straightforward, the integrals of $\mathrm{A}$ and $\mathrm{P}$ involve the singularity $G(\bm{r}',\bm{r}) \propto log(|\bm{r}'-\bm{r}|)$. For $\mathrm{P}$, this singularity only occurs for nodes on the boundary, for which we already know $\psi$. These nodes are simply skipped and rows in the matrix are replaced with rows of the identity matrix. For interior nodes, the singularity is only approached, but not reached, so an adaptive quadrature routine from the QUADPACK~\cite{quadpack} is sufficient.

For $\mathrm{A}$, while the integral itself is still convergent, the integration domain contains the logarithmic singularity. As a result, the fixed 6th-order quadrature approach of Crow~\cite{Crow1993} is used for overlapping edges in the discretized double line integral, while QUADPACK is used for the remaining segments. This approach limits convergence for pd$>$3, where the quadrature is no longer exact. Additionally, the convergence at high resolution is further limited by accuracy in computing $G(\bm{r}',\bm{r})$ in the limit $\bm{r}' \to \bm{r}$. At present, a numerical cutoff is used followed by an analytic extrapolation of the logarithmic dependence to machine precision. While this is not exact, it is found to exhibit sufficient accuracy for present applications (see section~\ref{sec:ver_analytic_coil}). It is also worth noting that this singularity exists with the original outer-iteration-based Lackner-Von Hagenow method as well, and many existing codes only capture this to modest accuracy through fixed, low-order quarature schemes (eg. Romberg integration)~\cite{FreeGS}.

An alternative approach, introduced by Albanase, Blum, de Barbieri, and employed by the FEEQS~\cite{Heumann2015} and NICE~\cite{Faugeras2020} codes utilizes an analytic result for semi-circular domains~\cite{albanese1986}. However, it is not always desirable to require a semi-circular outer domain, for example if one wishes to model only the region inside the vacuum vessel or some other internal structure. While this is not a particularly harsh limitation our approach alleviates this issue with minimal additional effort.

\section{Code description}
\label{sec:code_desc}
TokaMaker is designed to solve the model presented in eqs.~\ref{eq:grad_shaf_domains} and \ref{eq:grad_shaf_full} in two general formulations with many possible configurations of each corresponding to different sets of known or desired quantities. In this section, we provide a brief description of each of these modes, along with options and methods unique to each of these operations. Full descriptions of this capability, along with examples, are included on the project GitHub at \href{https://github.com/hansec/OpenFUSIONToolkit}{https://github.com/hansec/OpenFUSIONToolkit}.

\subsection{Single-point equilibria}
When designing a new device or discharge, it is desirable to compute one or more target equilibria, which are generally defined in terms of a subset of quantities like the desired shape, plasma current ($I_p = \int_{\mathcal{P}} J_{\phi} dA$), confined pressure, which is often expressed as $\beta = \int \frac{2 \mu_0 P}{B^2} dV$, plasma current and pressure profiles, currents in equilibrium field coils, and possibly other parameters. TokaMaker supports a variety of configurations in this capacity, all of which utilize a fixed-point iteration to handle the nonlinearity in eq.~\ref{eq:grad_shaf_disc}. In the most common application, fixed shapes are specified for $F^2(\psi)$ and $P(\psi)$, with unknown scale factors $\alpha_{F}$ and $\alpha_{P}$ applied to each profile respectively. At each step, the plasma contribution to $\psi$ is then computed as
\begin{equation} \label{eq:inv_step}
\psi^{n+1}_{I_p} = \alpha^{n+1}_F \psi^{n+1}_{F} + \alpha^{n+1}_P \psi^{n+1}_{P},
\end{equation}
where $\Delta^* \psi^{n+1}_{F} = -\frac{1}{2 \alpha^{n}_F}\frac{\partial F^2(\psi^{n})}{\partial \psi}$ and $\Delta^* \psi^{n+1}_{P} = - \frac{\mu_0 R^2}{\alpha^{n}_P} \frac{\partial P(\psi^{n})}{\partial \psi}$. $\alpha_F$ and $\alpha_P$ are determined by the solution to a 2x2 system of equations for targets and their linearized dependence on $\alpha_i$, which is just the identity matrix if $\alpha_{F}$ and $\alpha_{P}$ are fixed. A more frequent case is with targets for plasma current ($\bar{I}_{P}$) and radial location of the magnetic axis ($\bar{R}_{0}$) resulting in the system
\begin{equation} \label{eq:inv_ls}
\begin{bmatrix}
\bar{I}_{P} \\
-\nabla_R \psi^{n}_v (\bar{R}_{0},Z_0)
\end{bmatrix}
=
\begin{bmatrix}
-\frac{1}{2 \alpha^{n}_F} \int \frac{\partial F^2(\psi^{n})}{\partial \psi} dA & - \frac{\mu_0 R^2}{\alpha^{n}_P} \int \frac{\partial P(\psi^{n})}{\partial \psi} dA \\
\nabla_R \psi^{n+1}_{F} (\bar{R}_{0},Z_0) & \nabla_R \psi^{n+1}_{P} (\bar{R}_{0},Z_0) 
\end{bmatrix}
\begin{bmatrix}
\alpha_F \\
\alpha_P
\end{bmatrix},
\end{equation}
where $\psi_v$ is the solution to eq.~\ref{eq:grad_shaf_disc} in $\mathcal{S}$ and $\mathcal{C}$. Alternate constraints for row 2 include target pressure on axis, stored energy, and the ratio $I_{P,F}/I_{P,P} \approx \frac{1}{\beta_p} - 1$, while constraints for row 1 are presently limited to $I_P$ and $\alpha_{F}$.

If a plasma shape is specified but the coil currents are unknown, a so-called \textit{Inverse Equilibrium} calculation, then their values are next updated to minimize the least-square error between a set of isoflux targets, $\min_{I_{\mathcal{C}_i}} \left\{ \psi(\bm{r}_{i,iso}) - \psi(\bm{r}_{ref}) \right\}$, and saddle targets, $\min_{I_{\mathcal{C}_i}} \left\{ |\nabla \psi(\bm{r}_{i,saddle})| \right\}$, subject to a regularization matrix ($\mathrm{R}$) with targets ($\bar{I}_{\mathcal{C}_i}$) and optional bounds ($I^0_{\mathcal{C}_i} \leq I_{\mathcal{C}_i} \geq I^1_{\mathcal{C}_i}$)
\begin{equation} \label{eq:coil_ls}
\begin{bmatrix}
\psi_{\mathcal{P}}(\bm{r}_{ref}) + \psi_{\mathcal{S}}(\bm{r}_{ref}) \\
\vdots \\
-\nabla_R \left( \psi_{\mathcal{P}}(\bm{r}_{i,saddle}) + \psi_{\mathcal{S}}(\bm{r}_{i,saddle}) \right) \\
-\nabla_Z \left( \psi_{\mathcal{P}}(\bm{r}_{i,saddle}) + \psi_{\mathcal{S}}(\bm{r}_{i,saddle}) \right) \\
\vdots \\
\bar{I}_{\mathcal{C}_i} \\
\vdots
\end{bmatrix}
=
\begin{bmatrix}
\psi_{\mathcal{C}_j}(\bm{r}_{i,iso}) \dots \\
\vdots \\
\nabla_R \psi_{\mathcal{C}_j}(\bm{r}_{i,saddle}) \dots \\
\nabla_Z \psi_{\mathcal{C}_j}(\bm{r}_{i,saddle}) \dots \\
\vdots \\
\mathrm{R}_{i,j} \dots \\
\vdots
\end{bmatrix}
\begin{bmatrix}
I_{\mathcal{C}_j} \\
\vdots
\end{bmatrix}.
\end{equation}
Each row in the resulting system can be individually weighted to adjust the impact on the least-square error. Additionally, isoflux targets can be weighted by $1/|\nabla \psi(\bm{r}_{i,iso})|$ at the current iteration to weight each point's error more evenly in physical space. If bounds are employed, the BVLS~\cite{Lawson1995} is used, otherwise the standard least-squares approach is used. Additional constraints on coils, such as equal currents coils arranged in series, can also be applied through the regularization matrix. Note that $\psi_{\mathcal{C}_i}$ can be precomputed as the spatial variation in coil current and is fixed so only the scale factor is updated on each iteration.
\begin{figure}[h]
\begin{center}
    \includegraphics[width=0.4\textwidth]{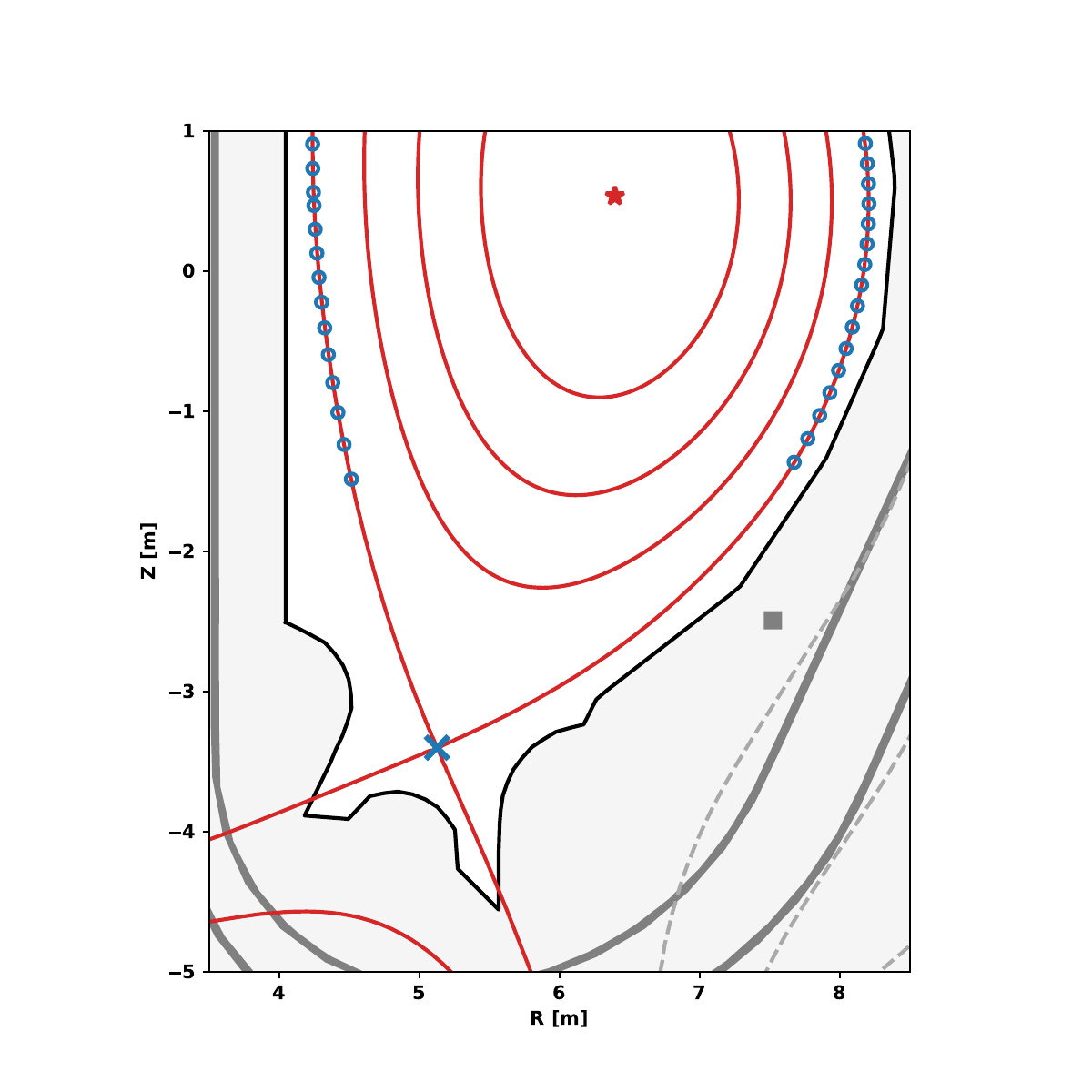} 
    \caption{Constraints (blue) applied, focusing on the X-point, to produce the equilibrium shown in figure~\ref{fig:ITER_eq_ex}. With an appropriate coilset, the resulting X-points (crosses) and isoflux points (circles) can be well matched by the final equilibrium (red).}
    \label{fig:ITER_con_ex}
\end{center}
\end{figure}

Even if coil currents are specified, a so-called \textit{Forward Equilibrium} calculation, it is often necessary to allow some flexibility and a target vertical position due to physical instability that exists in certain equilibria that will also manifest within the nonlinear solve. To alleviate this, a \textit{Vertical Stability Coil} (VSC), which is usually formed by a pair of coils with equal and opposite currents, can be specified along with a target for the vertical position of the magnetic axis ($\bar{Z}_0$). This coil may be a real coil used for similar control in an actual device or an artificial coil solely used for the numerical calculation. When active, the constraint
\begin{equation} \label{eq:vsc_con}
-\nabla_Z \left(\psi_{\mathcal{P}} (R_{0},\bar{Z}_0) + \psi_{\mathcal{S}} (R_{0},\bar{Z}_0) + \psi_{\mathcal{C}} (R_{0},\bar{Z}_0) \right) = \nabla_Z \psi_{VSC} (R_{0},\bar{Z}_0),
\end{equation}
which is similar to the last row in eq.~\ref{eq:inv_ls}, is used to determine the VSC current.

By default, the time-dependent term in eq.~\ref{eq:grad_shaf_disc} is omitted, consistent with an infinte-time equilibrium. However, quasi-static equilibria with self-consistent eddy currents in passive conductors can also be computed by setting a reference flux $\psi_{t - \Delta t}$, $\Delta t$, and $\eta$ in each region of $\mathcal{S}$. In this case
\begin{equation} \label{eq:inv_eddy}
\left( \int \frac{1}{R} \nabla v \cdot \nabla u \, dA - \int_{\mathcal{S}} \frac{v}{\eta R} \frac{u}{\Delta t} \, dA \right) \psi^{n+1}_{\mathcal{S}} = \int_{\mathcal{S}} \frac{v}{\eta R} \frac{\psi^{n+1}_{\mathcal{P}} + \psi^{n+1}_{\mathcal{C}} -\psi_{t-\Delta t}}{\Delta t} \, dA
\end{equation}
is solved on each iteration after the prior two steps to update the contribution of currents in $\mathcal{S}$ to the solution.

Fixed boundary equilibria can also be computed using this approach where only a plasma region is present with Dirichelt BCs $\psi_b = 0$. This renders many of the steps described above unnecessary, although the plasma updates in eq.~\ref{eq:inv_step} and associated constraints are still applied. Functionality is also included in TokaMaker to compute the vacuum flux, which must be provided by coils, required to reproduce the plasma boundary for a given fixed boundary equilibrium. This allows optimization of the placement and design of coils for target equilibria without the need to compute many corresponding free-boundary equilibria, which may be poorly behaved during intermediate optimization steps. Further optimization can then be performed with free-boundary equilibrium calculations to finalize the integrated system.

\subsection{Time-dependent equilibria}
For time-dependent equilibrium cases, the eddy currents in structural regions $\mathcal{S}$ play a significant role in the calculation -- often providing the current necessary for force balance. In general, there are two types of time-dependent calculations that are of interest: 1) Extraction of a linearized model of the time-dependent dynamics of the system and 2) Full nonlinear simulations of the evolution of plasma equilibria during natural and actively controlled evolution.

\noindent\underline{\textit{Linear}}

Linearizing eq.~\ref{eq:grad_shaf_disc} about an equilibrium $\psi_0$ with respect to a perturbation $\Delta \psi$ the system becomes
\begin{equation} \label{eq:time_dep_linear}
\begin{split}
    0 &= \int \frac{1}{R} \nabla v \cdot \nabla \left( \Delta \psi \right) \, dA - \int_{\mathcal{S}} \frac{v}{\eta R} \frac{\Delta \psi}{\Delta t} \, dA \\
    &+ \int_{\mathcal{P}} v \left( \frac{1}{2 R}\frac{\partial^2 F^2(\psi_0)}{\partial \psi^2} + \mu_0 R \frac{\partial^2 P(\psi_0)}{\partial \psi^2} \right) \Delta \psi  \, dA  \\
    &- \int_{\mathcal{P}} v (1-\hat{\psi}_0) \left( \frac{1}{2 R}\frac{\partial^2 F^2(\psi_0)}{\partial \psi^2} + \mu_0 R \frac{\partial^2 P(\psi_0)}{\partial \psi^2} \right) \Delta \psi(\bm{r}_{lim}) \, dA \\
    &- \int_{\mathcal{P}} v \hat{\psi}_0 \left( \frac{1}{2 R}\frac{\partial^2 F^2(\psi_0)}{\partial \psi^2} + \mu_0 R \frac{\partial^2 P(\psi_0)}{\partial \psi^2} \right) \Delta \psi(\bm{r}_{O}) \, dA,
\end{split}
\end{equation}
where the last two integrals are only included if $F^2$ and $P$ are defined in terms of the normalized coordinate $\hat{\psi} = [\psi - \psi(\bm{r}_O)]/[\psi(\bm{r}_{lim}) - \psi(\bm{r}_O)]$, where $\bm{r}_{lim}$ and $\bm{r}_O$ are the position of the limiter point (contact or X-point) and magnetic axis respectively. It is worth noting that the linearization can only capture variation with respect to a single limiting point, so care must be taken to interpret results if the last two terms are included and the equilibrium has two X-points with very similar values of $\psi$ as in so-called double null configurations.

This system can be assembled and used to study linear dynamics for control or other applications, but often it is of interest to perform eigenvalue analysis to determine if unstable modes exist in the system and their structure. In this case we can recast eq.~\ref{eq:time_dep_linear} as
\begin{equation} \label{eq:time_dep_linear_eig}
\begin{split}
    \omega \int_{\mathcal{S}} \frac{1}{\eta R} v \Delta \psi \, dA &= \int \frac{1}{R} \nabla v \cdot \nabla \left( \Delta \psi \right) \, dA \\
    &+ \int_{\mathcal{P}} v \left( \frac{1}{2 R}\frac{\partial^2 F^2(\psi_0)}{\partial \psi^2} + \mu_0 R \frac{\partial^2 P(\psi_0)}{\partial \psi^2} \right) \Delta \psi  \, dA  \\
    &- \int_{\mathcal{P}} v (1-\hat{\psi}_0) \left( \frac{1}{2 R}\frac{\partial^2 F^2(\psi_0)}{\partial \psi^2} + \mu_0 R \frac{\partial^2 P(\psi_0)}{\partial \psi^2} \right) \Delta \psi(\bm{r}_{lim}) \, dA \\
    &- \int_{\mathcal{P}} v \hat{\psi}_0 \left( \frac{1}{2 R}\frac{\partial^2 F^2(\psi_0)}{\partial \psi^2} + \mu_0 R \frac{\partial^2 P(\psi_0)}{\partial \psi^2} \right) \Delta \psi(\bm{r}_{O}) \, dA,
\end{split}
\end{equation}
where $\omega$ is a continuous replacement for $1/\Delta t$. The eigenspectrum of this system can then be computed by inverting the RHS and solving the system
\begin{equation} \label{eq:time_dep_linear_eig2}
     \mathrm{RHS}^{-1} \int_{\mathcal{S}} \frac{1}{\eta R} v \Delta \psi \, dA = \lambda \Delta \psi,
\end{equation}
for the eigenvalues $\lambda = 1/\omega$. While this can be done with a direct approach, we are often only interested in a few of the most unstable and/or stable modes of the system. So, TokaMaker instead employs an iterative Arnoldi method through the ARPACK library with a shift to compute the eigenvalues closest to some expected fastest growth rate ($\omega_s = -\gamma_{expected}$). This expected growth rate is manually specified, but can be readily approximated as a few times the slowest decay time ($\tau_{L/R}$) of the wall, which can be computed using the same method and setting the plasma terms to zero.
\begin{figure}[h]
\begin{center}
    \includegraphics[width=0.4\textwidth]{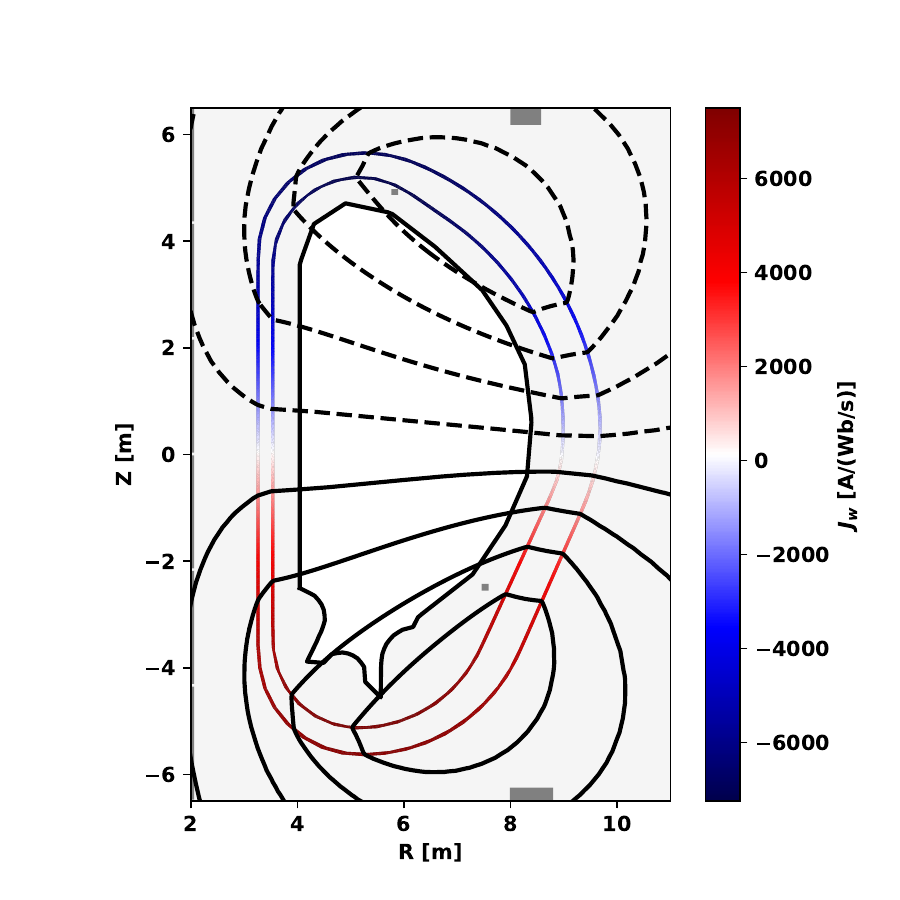} 
    \caption{Up-down asymmetric decay mode for eddy currents in ITER with a uniform $\eta = 6.9 \times 10^{-7}$~$\Omega$-m in the inner and outer vacuum vessel ($\tau_{L/R} = 337$~ms).}
    \label{fig:ITER_decay_ex}
\end{center}
\end{figure}

\noindent\underline{\textit{Nonlinear}}

For the nonlinear case, the system of equations is the same as in the equilibrium case above, but we use a more robust nonlinear solution method based on a matrix-free Newton-Krylov approach. While this is not strictly necessary for many of the cases considered here, it leads to better performance and provides greater flexibility for the addition of new capabilities, such as coupled evolution of $\psi$, $F$, and $P$ consistent with physical transport models. The linear system within the MFNK method is solved using FGMRES that is preconditioned using an LU factorization of an explicitly formed approximate Jacobian. As the objective of this type of simulation is generally to observe and study the evolution of equilibria, the change in solution from timestep to timestep is usually modest. As a result, the vacuum operator
\begin{equation} \label{eq:time_dep_vac}
    \mathcal{J} \approx \int \frac{1}{R} \nabla v \cdot \nabla u \, dA - \int_{\mathcal{S}} \frac{v}{\eta R} \frac{u}{\Delta t} \, dA
\end{equation}
often works well as an approximate Jacobian providing robust and rapid convergence. If necessary, additional terms in the Jacobian, as expressed in eq.~\ref{eq:time_dep_linear} can also be included.

On each timestep the functions $F$ and $P$, coil currents (feedforward or feedback), and targets for $I_p$, or $\alpha_F$, and  $I_{P,F}/I_{P,P}$, or $\alpha_P$, can be adjusted. At present, the evolution of these parameters must be specified by the user, but future work will focus on implementing transport equations to allow the self-consistent evolution of these and other quantities.

\section{Verification tests}
\label{sec:verification}
To verify the methods described above and their implementation in TokaMaker, a series of benchmarks were performed. Both analytic test cases for fixed and free-boundary cases were studied. Comparison to existing community tools for a range of practical cases were also performed, which is partly due to the lack of analytic cases for full free-boundary equilibria with vacuum regions, where $\frac{\partial F^2}{\partial \psi} = \frac{\partial P}{\partial \psi} = 0$, and the desire to validate performance on realistic/complex configurations. Beyond the tests presented here, the underlying code used in TokaMaker, which was previously used in PSI-Tri~\cite{HansenThesis,Hansen2017}, has been applied previously to a variety of problems in both tokamaks~\cite{Hansen2017} and spheromaks~\cite{Sutherland2021}.

\subsection{Analytic verification}
\label{sec:ver_analytic}
In this section, we compare the results for TokaMaker to a set of well-known analytic solutions to eq.~\ref{eq:grad_shaf} for different mesh resolutions and basis function polynomial degree (pd). In practice, it is expected that the tool will be used mostly near the low resolution, low order (pd=2) end of these studies as high-accuracy is not required for most such applications. However, it is still useful to perform such convergence studies to demonstrate the efficacy and robustness of the tool. Additionally, there may be certain areas where improved accuracy over existing tools would be beneficial (eg. stability)~\cite{Glasser2016,Snyder2002,Xing2021}.

\subsubsection{Spheromak}
\label{sec:ver_analytic_sph}
\begin{figure}[h]
\begin{center}
    \includegraphics[width=0.8\textwidth]{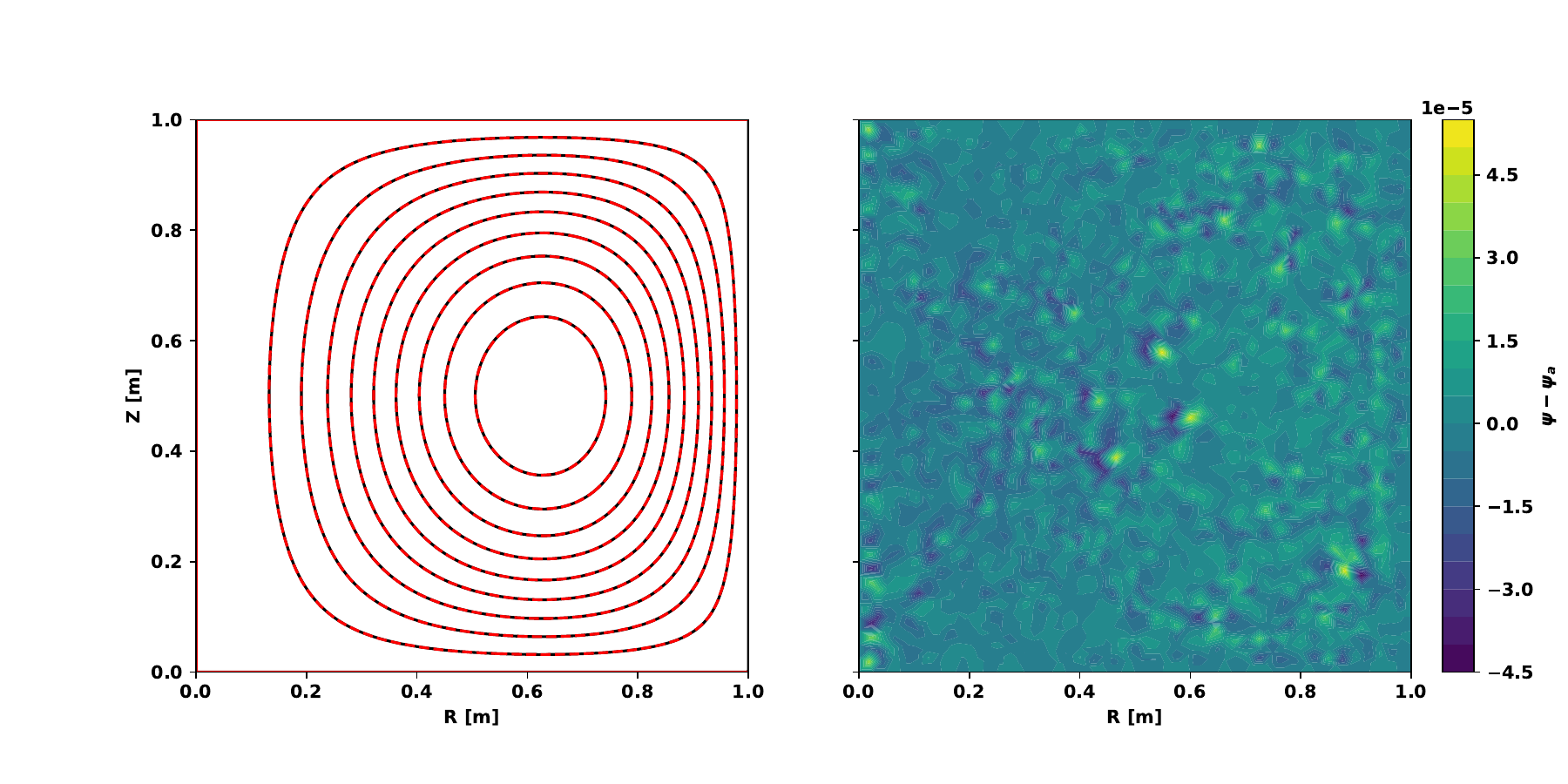} 
    \caption{Example results from the spheromak test case showing (left) comparison of flux surfaces from TokaMaker (red) and analytic solution (black) for pd=2 and $\Delta x = 0.05$ and (right) local error across the grid for the same solution.}
    \label{fig:sph_conv_setup}
\end{center}
\end{figure}

For a plasma whose boundary is specified and includes the geometric axis, with no externally applied magnetic fields (toroidal or poloidal), zero-$\beta$ ($\frac{\partial P}{\partial \psi} = 0$), and $\frac{\partial F^2}{\partial \psi} \propto \psi$ eq.~\ref{eq:grad_shaf} becomes the eigenvalue equation
\begin{equation}\label{eq:gs_spheromak}
\nabla^{*} \psi = \lambda^2 \psi.
\end{equation}
The lowest-eigenvalue solution to this equation is the spheromak~\cite{BellanBook,Rosenbluth1979}, which in a rectangular domain of height $h$ and radius $a$ can be expressed analytically as
\begin{equation}\label{eq:spheromak_analytic}
\psi(\bm{r}) = \psi_0 \frac{\gamma_{1,1} R}{\chi_{0,1}} \frac{J_1(\gamma_{1,1} R)}{J_1(\chi_{0,1})},
\end{equation}
where $\gamma_{1,1} = \chi_{1,1}/a$, and $\chi_{i,1}$ is the first zero of the i-th Bessel function $J_i$. Figure~\ref{fig:sph_conv_setup} shows a comparison of the resulting flux surfaces and error in $\psi$ for the analytic and TokaMaker solutions at a modest resolution ($\Delta x = 0.05$) of about 20 cells per direction ($\approx$~2,500 DOF), showing good agreement with the analytic solution.

\begin{figure}[h]
\begin{center}
    \includegraphics[width=0.4\textwidth]{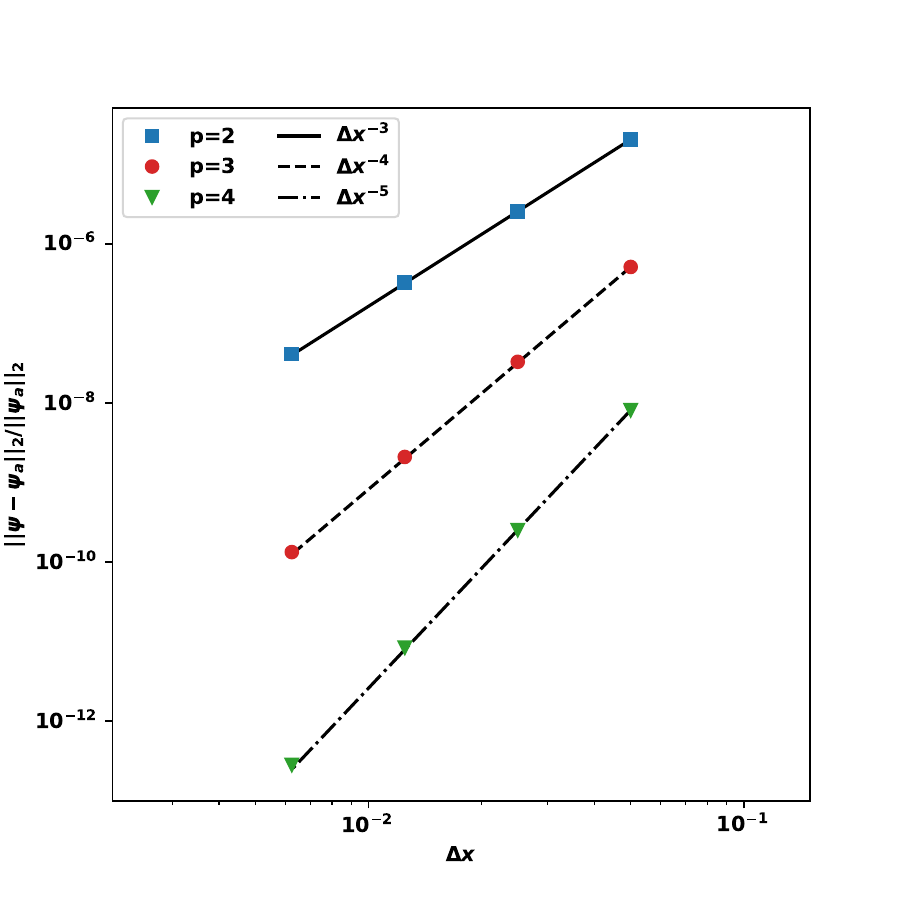} 
    \caption{Convergence of the error between the TokaMaker and analytic solutions for the spheromak test case. For each polynomial order, the results (markers) match the expected convergence rates (lines).}
    \label{fig:sph_conv}
\end{center}
\end{figure}
As spatial resolution and/or polynomial degree are increased, the error in the solution converges to zero at the expected rates of O($\Delta x^{-(pd+1)}$), as shown in figure~\ref{fig:sph_conv}.

\subsubsection{Solov'ev}
\label{sec:ver_analytic_solo}
\begin{figure}[h]
\begin{center}
    \includegraphics[width=0.8\textwidth]{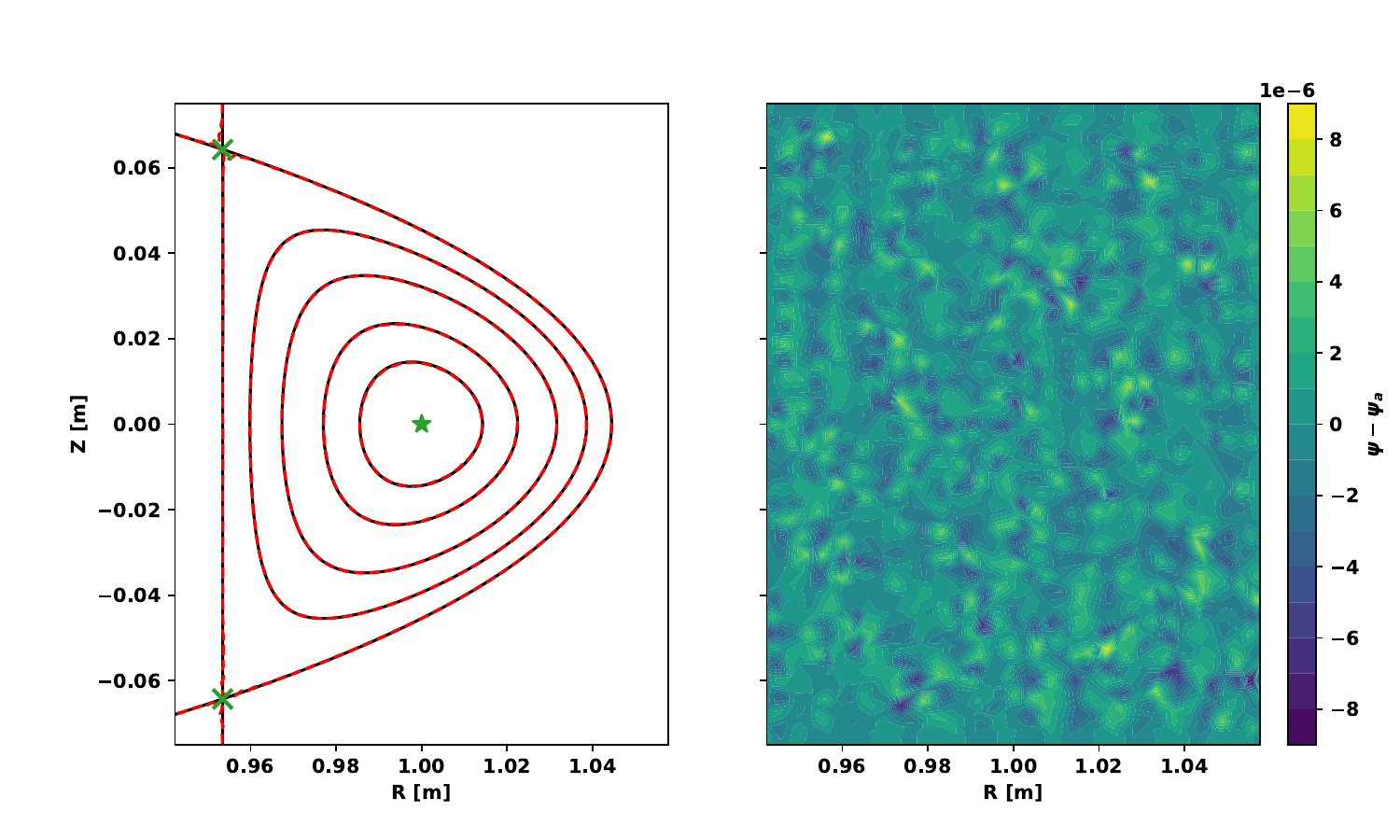} 
    \caption{Example results from the Solov'ev test case showing a comparison of flux surfaces from TokaMaker (red) and the analytic solution (black) for pd=2 and $\Delta x = 0.0075$ (left) and local error across the grid for the same solution (right). X-points (saddle points) and the magnetic axis (local maximum) are shown with green markers.}
    \label{fig:solo_conv_setup}
\end{center}
\end{figure}

The Solov'ev equilibria~\cite{Solovev1968} are a set of analytic solutions to eq.~\ref{eq:grad_shaf} with an externally applied toroidal field that have been extensively used for analytic treatment of tokamaks~\cite{Zheng1996,Xu2019,Cerfon2010} as well as benchmarking G-S codes~\cite{Howell2014}. For $\frac{\partial P}{\partial \psi} = -a$ and $\frac{\partial F^2}{\partial \psi} = \frac{-b}{2} R_0^2$, the solution to eq.~\ref{eq:grad_shaf} is a quartic function of the form
\begin{equation}\label{eq:gs_solovev}
\psi(\bm{r}) = \frac{1}{2} (b+c_0) R_0^2 Z^2 + c_0 R_0 \zeta Z^2 + \frac{1}{2} (a-c_0) R_0^2 \zeta^2,
\end{equation}
where $c_0$ is an additional constant and $\zeta = \frac{R^2 - R_0^2}{2R_0}$.

As a benchmark, we compare the case $R_0=1.0$, $a=1.2$, $b=-1.0$, and $c_0 = 1.1$ from \cite{Xu2019}, which yields an up-down symmetric equilibrium with two X-points (saddles), where the poloidal field goes to zero. Figure~\ref{fig:solo_conv_setup} shows a comparison of the resulting flux surfaces and error in $\psi$ for the analytic and TokaMaker solutions at a modest resolution ($\Delta x = 0.0075$) of about 20 cells in the vertical direction ($\approx$~2,000 DOF), showing good agreement with the analytic solution.

Note that in the Solov'ev solutions $\frac{\partial P}{\partial \psi}$, $\frac{\partial F^2}{\partial \psi}$, and $J_{\phi}$ are all non-zero everywhere in space, which is non-physical for a real system. Indeed, while the solution may appear to be free-boundary, it is not in the sense of interest for application as a Dirichlet BC is applied using the analytic $\psi$ from eq.~\ref{eq:gs_solovev}.

\begin{figure}[h]
\begin{center}
    \includegraphics[width=0.4\textwidth]{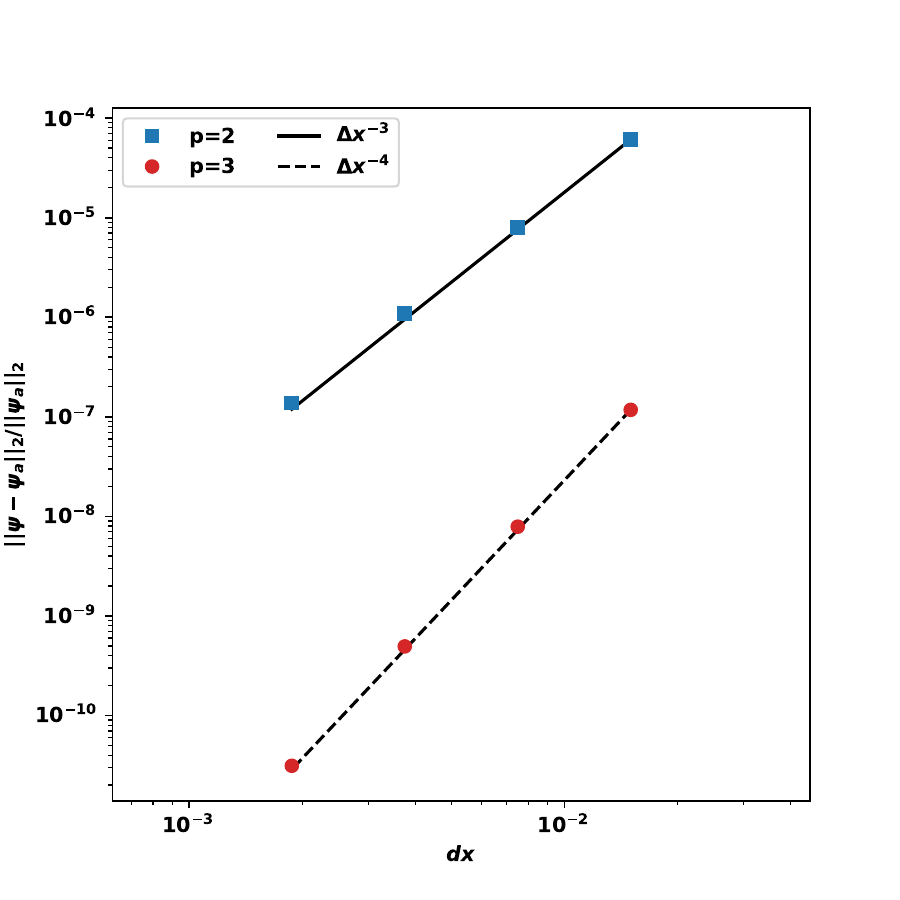}
    \includegraphics[width=0.4\textwidth]{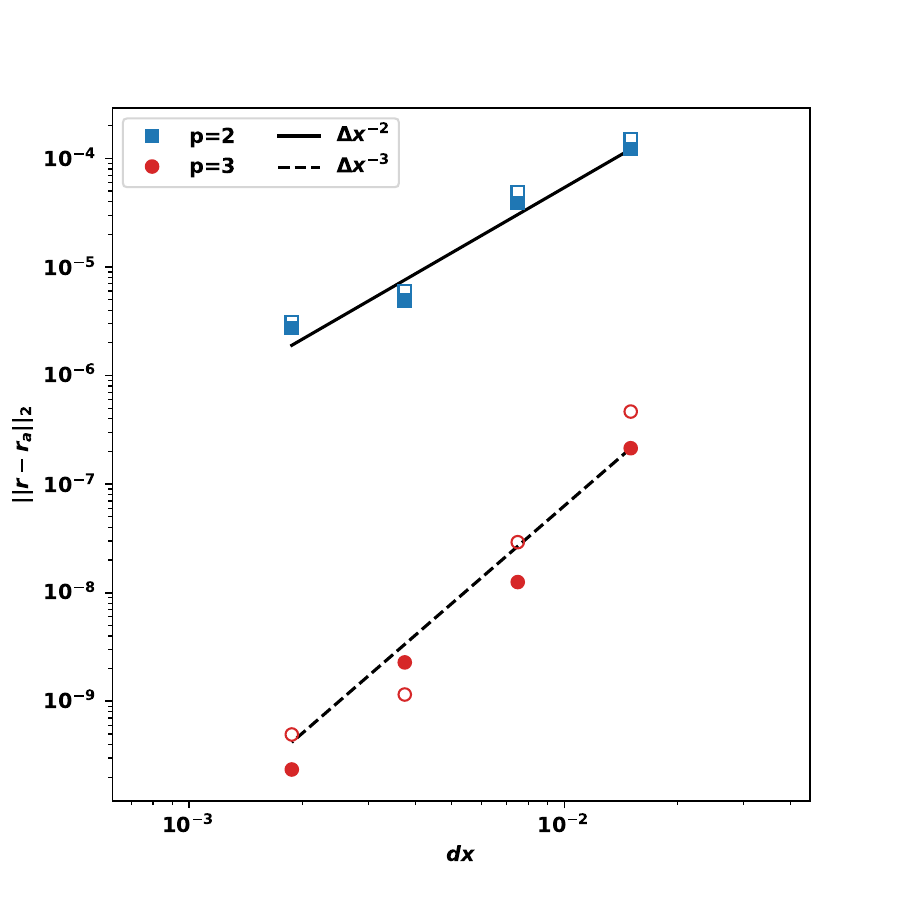} 
    \caption{Convergence of the error between the TokaMaker and analytic solutions for the Solov'ev test case for left: $\psi$ and right: the location of the X-points (open) and O-points (filled). For each polynomial order, the results (markers) match the expected convergence rates (lines). Note that the Solov'ev test case is exact at 4th order, so results for $p>3$ are not shown.}
    \label{fig:solo_conv}
\end{center}
\end{figure}
As spatial resolution and/or polynomial degree are increased, the error in the solution converges to zero at the expected rates of O($\Delta x^{-(pd+1)}$), as shown in figure~\ref{fig:solo_conv}. This benchmark is also used to verify the location of O-points (local maxima) and X-points (saddles) in TokaMaker, which are used to define the plasma extent in both physical and $\psi$ space. The method used by TokaMaker is able to correctly identify these points to sub-gridscale. Note that results are only shown up to cubic basis functions, as the analytic solution is quartic, so representations of pd$>$3 are accurate to machine precision.

\subsubsection{Vacuum coil}
\label{sec:ver_analytic_coil}
\begin{figure}[h]
\begin{center}
    \includegraphics[width=0.8\textwidth]{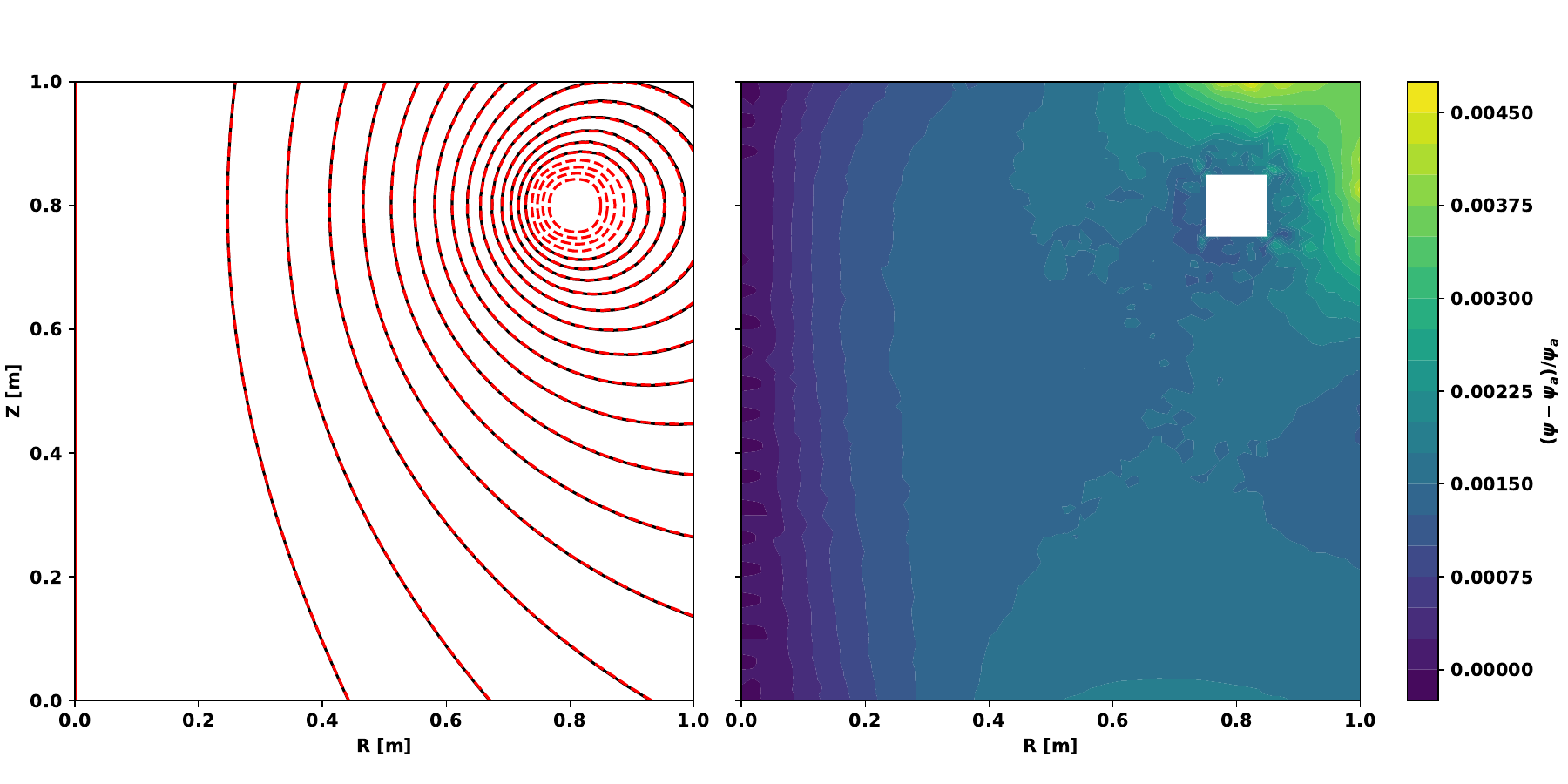} 
    \caption{Example results from the free-boundary coil test case showing (left) a comparison of flux surfaces from TokaMaker (red) and analytic solution (black) for pd=2 and $\Delta x = 0.05$ and (right) local error across the grid for the same solution. The analytic solution is only computed outside the coil to avoid integrating the singularity in the axisymmetric Green's function.}
    \label{fig:coil_conv_setup}
\end{center}
\end{figure}

The previous two test cases verified the ability of TokaMaker to solve the elliptic PDE with Dirichlet BCs and identify important features of $\psi$ such as X- and O-points. For free-boundary cases, we must also verify that the boundary condition described in section~\ref{sec:boundary_conditions} are also implemented correctly. However, there are no analytic solutions for true free-boundary equilibria with zero current beyond the LCFS. Instead, we will use the vacuum solution to current in a square toroidal conductor of constant current density as an analytic test case. In practice, such a solution can be computed efficiently using a Dirichlet BC with nodal boundary values computed from integration of the analytic Green's function. This makes the free-boundary BC unnecessary, but it nonetheless makes a useful benchmark for the free-boundary BC for other current distributions (eg. plasma currents) that are not fixed in time.

For this test case, $\psi$ produced by a uniformly distributed current flowing in a 0.1 x 0.1 m square cross-section coil centered at R=0.75, Z=0.75 is computed. Tokamaker uses a 1.0 x 1.0 m domain (0.2 m separation in R and Z between the coil and the mesh boundary). The analytic solution is computed by integrating the Green's function from above over the poloidal cross-section of the coil using ODEPACK. Figure~\ref{fig:coil_conv_setup} shows a comparison of the resulting flux surfaces and error in $\psi$ for the analytic and TokaMaker solutions at a modest resolution ($\Delta x = 0.05$) of about 20 cells in each direction in the vacuum region and a higher resolution ($\Delta x = 0.01$) in the coil in order to allow convergence studies without changing the resolution in the coil itself.

\begin{figure}[h]
\begin{center}
    \includegraphics[width=0.4\textwidth]{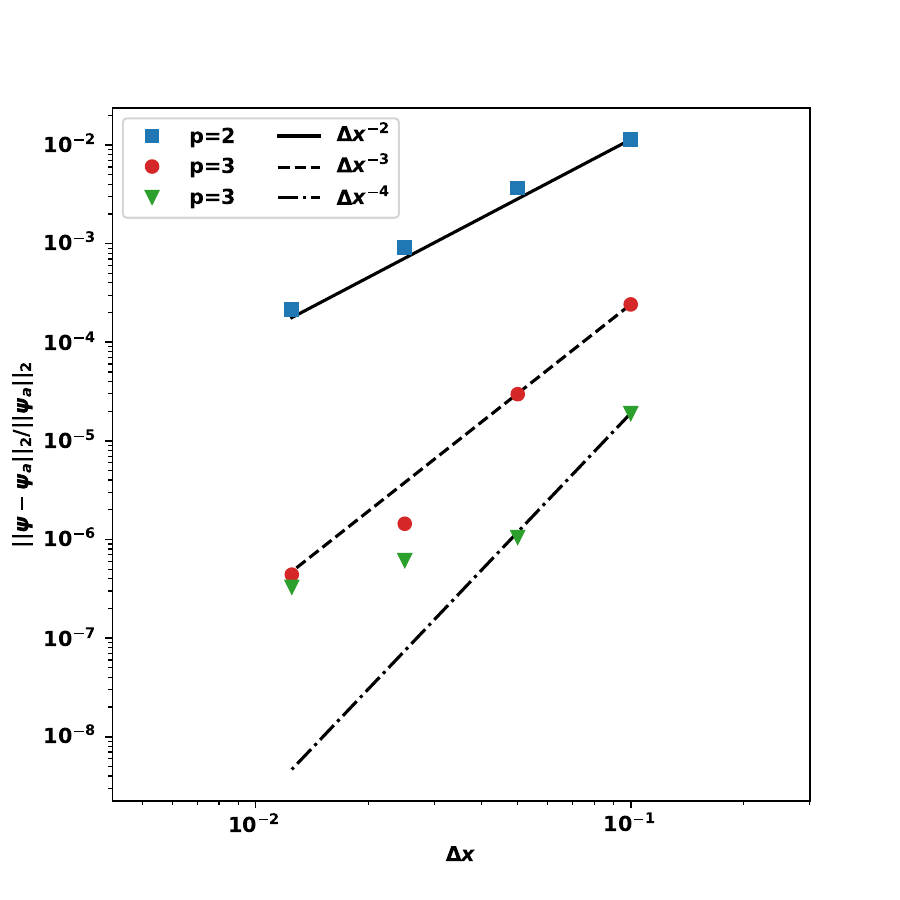} 
    \caption{Convergence of the boundary flux error between the TokaMaker and analytic solutions for the free-boundary coil test case. Convergence at low error and high polynomial order is limited by accuracy of quadrature.}
    \label{fig:coil_conv}
\end{center}
\end{figure}
The convergence behavior for this test case is shown in figure~\ref{fig:solo_conv}. In contrast to prior cases, convergence is evaluated only using the nodes on the boundary, skipping those in the interior, to emphasize the effect of the BC. The resulting convergence exhibits more complex behavior than the prior two due to the difficulty in evaluating the integrals necessary for the boundary condition to high order. For pd$\leq$3, the error converges uniformly as the spatial resolution is increased. However, a floor is observed in the convergence of the error for pd=4, after which the error begins to decrease much more slowly. It is expected that the fixed-order accuracy of performing the integration, along with limitations on evaluating the Green's function in the vicinity of the singularity, is the cause of this floor. Future work will attempt to address this, but for the moment, the level of accuracy achievable is much higher than required for practical usage as shown in section~\ref{sec:ver_cross}.

\subsection{Cross-code verification}
\label{sec:ver_cross}
While analytic cases provide the ability to study convergence error to high accuracy, the lack of such cases for true free-boundary equilibria with physical coil geometries prevents such precise verification on realistic equilibria. In this section, we instead compare equilibria generated in two commonly used community tools with equilibria generated using TokaMaker with the same coil currents, flux profiles, and other relevant geometry. For these comparisons, we have chosen to use equilibria in the SPARC device~\cite{Creely2020}, primarily due to our group's recent experience applying detailed equilibrium analysis to this device.

\subsubsection{FreeGS}
\label{sec:ver_freegs}
\begin{figure}[h!]
    \centering
    \includegraphics[width=0.8\textwidth]{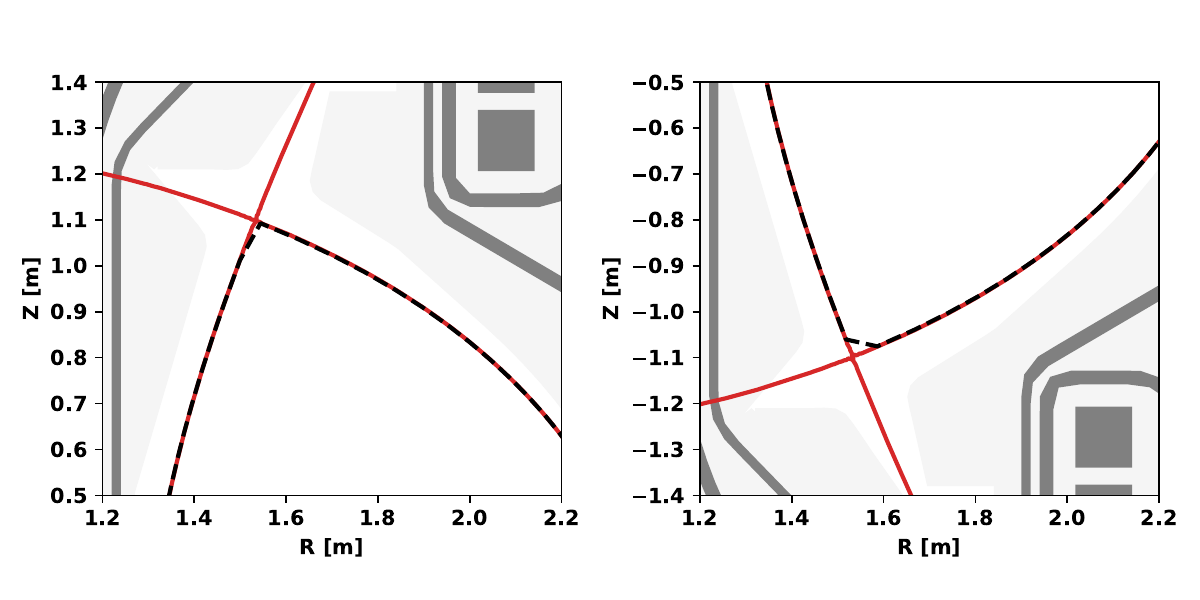}
    \caption{Comparison of the LCFS for TokaMaker (red contour) and FreeGS (black dashed contour) for matched coil currents and flux functions. Note the apparent deviation very near each X-point is an artifact of limited resolution in the equilibrium output file (gEQDSK) from FreeGS.}
    \label{fig:FGS_boundary_matches}
\end{figure}

FreeGS is an open-source, free-boundary Grad-Shafranov solver written in python~\cite{FreeGS}. Given a set of coils, plasma profiles and shape, FreeGS solves the inverse problem and determines coil currents, which produce a steady-state equilibrium solution using either the direct Biot-Savart approach or the Lackner and Von Hagenow method~\cite{Lackner1976,JardinBook} and an innner-outer iteration for boundary conditions. L-mode SPARC equilibria~\cite{Rodriguez2020} generated using FreeGS have been publically released by Commonwealth Fusion Systems~\cite{CFSGithub}, providing an ideal test case for TokaMaker and other new equilibrium codes.

For comparison, we computed an L-mode double null equilibrium in FreeGS using published $F*F'$ and $P'$ profiles~\cite{CFSGithub}. The same case was then computed in TokaMaker using a forward calculation with the same coil currents, flux profiles, and targets to match the $I_P$ and $P_0$ from the FreeGS calculation. The resulting equilibria match extremely well, as shown by the comparison of the TokaMaker and FreeGS last closed flux surfaces in figure~\ref{fig:FGS_boundary_matches}.

\subsubsection{EFIT}
\label{sec:ver_efit}
EFIT~\cite{Lao1985} is a G-S equilibrium generation and reconstruction code that is widely-used in the fusion community. Typically, EFIT is used to do so-called equilibrium reconstruction, which is a type of inverse problem where instead of shape constraints, diagnostics signals are used to constrain the resulting equilibrium. In this case, coil currents are typically known, but $F*F'$ and $P'$ including their global scales, which relate to $I_P$ and $\beta_P$, are unknown and must be fit.

\begin{figure}[h!]
    \centering
    \includegraphics[width=0.8\textwidth]{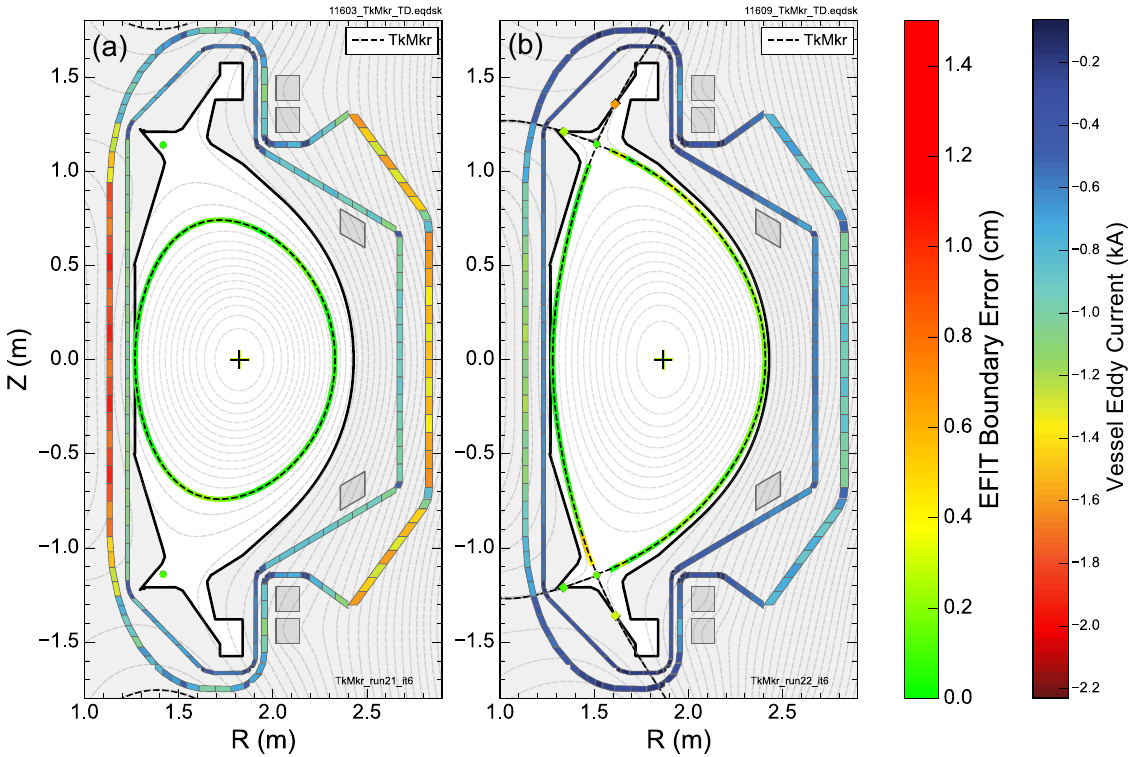}
    \caption{EFIT reconstruction boundary errors for (a) an inboard limited SPARC equilibrium and (b) a diverted, double null SPARC equilibrium. The eddy currents used in the EFIT reconstruction, which were derived from the time-dependent TokaMaker solutions, are also indicated. Note: the X-point and strike point locations are color coded with the same spatial bounary error shown in the color bar on the left.}
    \label{fig:EFIT_boundary_matches}
\end{figure}

While EFIT can also operate in a way similar to the cases described for TokaMaker above, we chose instead to use the reconstruction capability for comparison. For this benchmark, we take two equilibria out of a sequence of equilibria of a full example discharge in the SPARC tokamak. For this calculation, we include eddy currents in the calcultion with TokaMaker as described in sec.~\ref{sec:code_desc}. We then output the resulting equilibria using gEQDSK files and reconstruct the resulting equilibria in EFIT, including currents in the vacuum vessel structures. The resulting comparison, shown in figure~\ref{fig:EFIT_boundary_matches}, has excellent agreement between the EFIT and TokaMaker results at both times. While not shown in this comparison, the coil currents are also matched between the EFIT and TokaMaker equilibria as well.

\section{Conclusions and future work}
\label{sec:conclusions}
In this paper, we presented a new static and time-dependent MHD equilibrium code (TokaMaker) for axisymmetric configurations, based on the well known Grad-Shafranov equation. This code utilizes finite element methods on an unstructured triangular grid to enable capturing accurate machine geometry and simple mesh generation from engineering-like descriptions of present and future devices. The new code is designed for ease of use without sacrificing capability and speed through a combination of Python, Fortran, and C/C++ coding paradigms. TokaMaker is part of the broader Open FUSION Toolkit, which is fully open-source and available freely on GitHub (\href{https://github.com/hansec/OpenFUSIONToolkit}{https://github.com/hansec/OpenFUSIONToolkit}) including detailed documentation and examples.

We have presented a detailed description of the numerical methods of the code and validation of the implementation of those methods using both analytic test cases and cross-code validation with the FreeGS and EFIT codes. The results show expected convergence for polynomial orders 2-4 for fixed-boundary analytic test cases. Free-boundary convergence follows a similar trend, but is limited at very high accuracy due to limitations in performing the required integral in assembling the boundary conditions. As this limitation only occurs at error levels well below those sought in practice for these types of tools, it is not expected to impact application.

Future work includes further development of the time-dependent capabilities of the code, focusing on self-consistent evolution of internal $F*F'$ and $P'$ profiles subject to plasma transport. Additionally, improved boundary integration to remove the convergence limit observed in free-boundary cases at high tolerance will be pursued. Finally, we are also working on integration of this model within optimization frameworks for application to both scenario development and device design. 

\section*{Acknowledgements}
 This work was supported by the U.S. Department of Energy, Office of Science, Office of Fusion Energy Sciences under Award(s) DE-SC0019239, DE-SC0019479, DE-SC0022270, and DE-SC0022272. Cross-code verification on the SPARC tokamak was supported by Commonwealth Fusion Systems. C. Hansen was supported by DE-SC0019239 and DE-SC0019479. S. Guizzo, A.O. Nelson, and C. Paz-Soldan were supported by DE-SC0022270. S. Guizzo was also supported by Columbia University internal funds. M. Pharr was supported by DE-SC0022272. I.G. Stewart and D. Burgess were supported by Commonwealth Fusion Systems.

 The authors also would like the thank Holger Heumann for helpful discussions on time-dependent equilibria.

\vspace{16pt}
\small
\noindent Disclaimer: This report was prepared as an account of work sponsored by an agency of the United States Government. Neither the United States Government nor any agency thereof, nor any of their employees, makes any warranty, express or implied, or assumes any legal liability or responsibility for the accuracy, completeness, or usefulness of any information, apparatus, product, or process disclosed, or represents that its use would not infringe privately owned rights. Reference herein to any specific commercial product, process, or service by trade name, trademark, manufacturer, or otherwise does not necessarily constitute or imply its endorsement, recommendation, or favoring by the United States Government or any agency thereof. The views and opinions of authors expressed herein do not necessarily state or reflect those of the United States Government or any agency thereof.

\bibliographystyle{elsarticle-num} 
\bibliography{tmaker}





\end{document}